\documentclass[conference]{IEEEtran}
\usepackage{array}
\usepackage{graphicx}
\usepackage{afterpage} 
\usepackage{caption}
\usepackage{stfloats} 
\usepackage{float}
\usepackage{afterpage}
\usepackage{float}
\usepackage{placeins}

\usepackage{titlesec}
\titleformat{\section}{\large\bfseries}{\thesection}{1em}{}
\titleformat{\subsection}{\normalfont}{\thesubsection}{1em}{}
\titlespacing*{\section}{0pt}{12pt}{6pt}
\titlespacing*{\subsection}{0pt}{10pt}{4pt}
\renewcommand{\thesection}{\arabic{section}}
\renewcommand{\thesubsection}{\thesection.\arabic{subsection}}

\title{MIRAGE: Metadata-Integrated Repository Analysis and
Guided Enhancement for MSR Datasets}

\author{%
\centering
\setlength{\tabcolsep}{2pt}
\begin{tabular}{c c c}
\parbox[c]{0.21\linewidth}{\centering\bfseries\normalsize Aabia Ather} &
\parbox[c]{0.21\linewidth}{\centering\bfseries\normalsize Muhammad Usayd Ather} &
\parbox[c]{0.21\linewidth}{\centering\bfseries\normalsize Qurat-Ul-Ain Somroo} \\[0.2cm]
\parbox[c]{0.17\linewidth}{\centering\scriptsize SEECS, NUST\\Islamabad, Pakistan\\aather.msit25seecs@seecs.edu.pk} &
\parbox[c]{0.17\linewidth}{\centering\scriptsize SEECS, NUST\\Islamabad, Pakistan\\akhalid.msit25seecs@seecs.edu.pk} &
\parbox[c]{0.17\linewidth}{\centering\scriptsize SEECS, NUST\\Islamabad, Pakistan\\qsomroo.msit25seecs@seecs.edu.pk} \\[0.5cm]

\end{tabular}
}

\author{
\centering
\begin{tabular*}{\textwidth}{@{\extracolsep{\fill}}cccc@{}}
Aabia Ather & Muhammad Usayd Ather & Qurat-Ul-Ain Somroo & Muhammad Khuram Shahzad \\[4pt]
\footnotesize SEECS, NUST, Islamabad, Pakistan & \footnotesize SEECS, NUST, Islamabad, Pakistan & \footnotesize SEECS, NUST, Islamabad, Pakistan & \footnotesize SEECS, NUST, Islamabad, Pakistan \\[4pt]
\footnotesize aather.msit25seecs@seecs.edu.pk & \footnotesize maather.mscs25seecs@seecs.edu.pk & \footnotesize qsomroo.msit25seecs@seecs.edu.pk & \footnotesize mkhuram.shahzad@seecs.edu.pk
\end{tabular*}
}

\begin{document}

\maketitle

\begin{abstract}
This paper proposes an improved approach to the analysis of Mining Software
Repositories (MSR) datasets via metadata enrichment, FAIRness assessment, and topic-driven
approach. This research expands upon an earlier dataset directory created specifically
for the analysis of MSR datasets by adding new annotations to the datasets, enriching
the metadata categories, and offering more advanced filtering options. The metadata of
the MSR papers presented from 2013 to 2024 has been gathered using the Semantic
Scholar API. The analysis is based on Latent Dirichlet Allocation (LDA) topic modeling
and statistics. Dataset-level attributes were included into the expanded dataset
directory, namely, the repository hosting site, format, accessibility, reusability, and
dataset quality. It was revealed that the choice of repository hosting sites and data
format influences citation patterns and dataset usability. Moreover, the improved
annotation approach facilitated the analysis of the dataset directory and made it FAIR.
The study aims at increasing the reusability and searchability of the MSR datasets.
\end{abstract}
\vspace{-1em}

\section{Introduction}
The development of software platforms like GitHub and GitLab led to the creation of software engineering datasets. These datasets include source code repositories, issues, pull requests and developers interactions. They became very useful for researching Mining Software Repositories (MSR). This is because they help analyze software evolution developers activities and software quality.

For supporting research on MSR many dataset repositories and directories were created. However these datasets have some limitations.
To solve these problems, Diamantopoulos et al. Proposed the "Directory of MSR Datasets" project. This project provides a repository with metadata, citations and evaluation of datasets based on principles. FAIR stands for Findable, Accessible, Interoperable and Reusable data.The Mining Software Repositories (MSR) datasets help in analyzing the software evolution.

Despite its contributions the initial version of the Directory of MSR Datasets did not offer dataset annotation and filtration options. The Directory of MSR Datasets provides a repository containing metadata for the datasets.evaluation of datasets based on principles. The GitHub and GitLab platforms are software platforms.

Github :https://github.com/aabiaather/MIRAGE-Metadata-Integrated-Repository-Analysis-and-Guided-Enhancement-for-MSR-Datasets
\begin{figure}[htbp]
    \centering
    \includegraphics[width=0.47\textwidth]{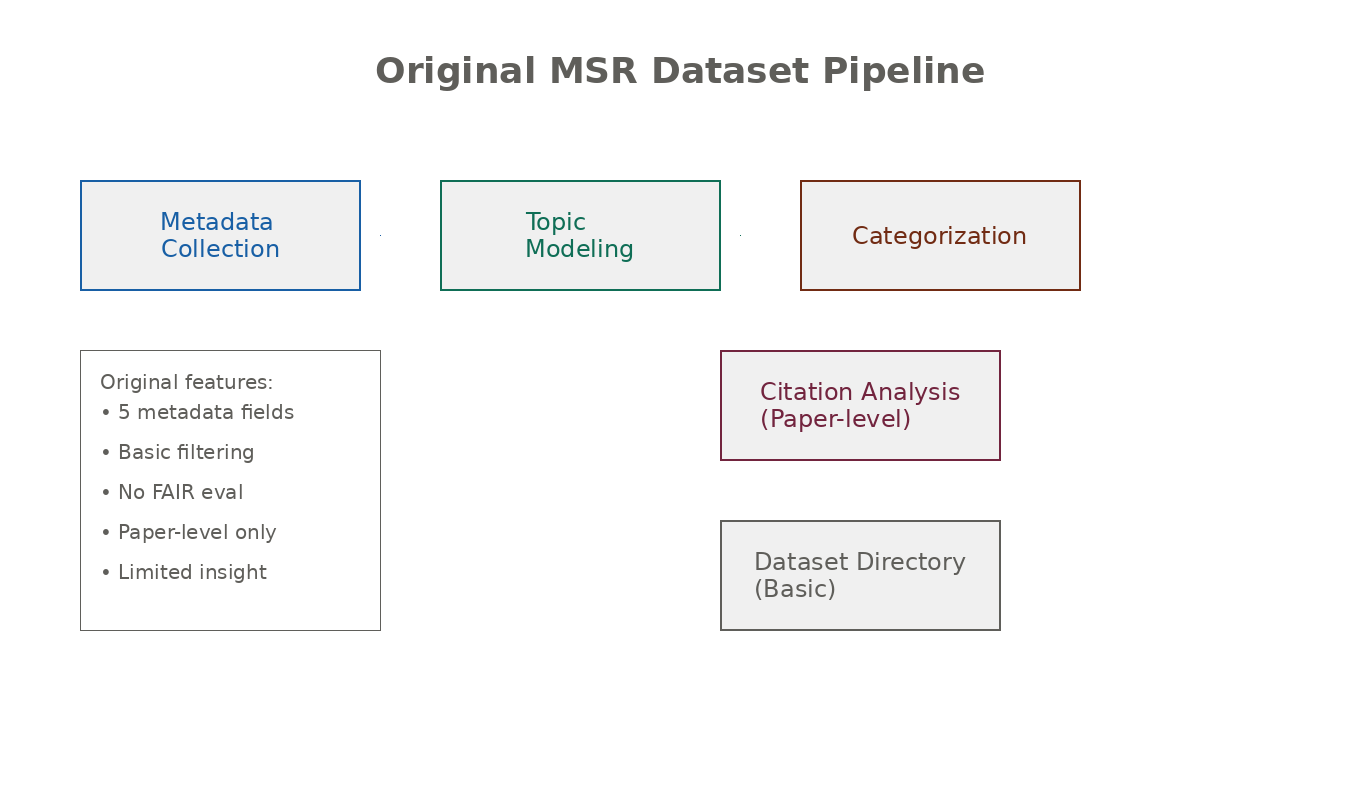}
    \label{fig:hosting}
\centering
\caption{Methodology of the original paper}
\end{figure}

In this paper, we aim at improving the initial idea and provide enriched metadata,
advanced dataset filtering, and dataset-level features extraction. We also analyze how
attributes of a dataset can influence citation counts and academic research impact.

\section{Related Work}

Several studies have focused on improving the usability and accessibility of Mining Software Repositories (MSR) datasets. Diamantopoulos et al. \cite{b1} proposed a directory of MSR datasets containing metadata, citation information, and FAIRness evaluation. Their work helped researchers locate and assess datasets more efficiently.

Kotti et al. \cite{b2} analyzed the usage patterns of MSR datasets and found that only a limited number of datasets receive significant attention from researchers. Their study highlighted the importance of dataset visibility and accessibility.

The PROMISE Repository \cite{b3} was one of the earliest collections of software engineering datasets for empirical research. Although useful, it lacked advanced organization, annotation, and filtering mechanisms.

The Public Git Archive \cite{b4} introduced a large-scale collection of GitHub repositories for source code analysis and machine learning research. However, the work mainly focused on dataset collection rather than improving dataset usability and metadata organization.

Similarly, GHTorrent \cite{b5} collected extensive GitHub activity data for software engineering research. Even though it became a widely used dataset source, it did not provide advanced filtering capabilities or FAIR-based dataset evaluation.

The FAIR Guiding Principles introduced by Wilkinson et al. \cite{b6} defined standards for making datasets Findable, Accessible, Interoperable, and Reusable. These principles are now widely used for evaluating scientific datasets and improving their reusability.

Other studies \cite{b7,b8} emphasized the importance of automated FAIRness assessment tools for improving dataset quality and research efficiency. These approaches help researchers evaluate datasets more effectively.

Several researchers \cite{b9} categorized MSR datasets into areas such as version control systems, software issues, and developer metrics. However, these studies did not provide enriched annotations or multidimensional metadata filtering techniques.

Our work extends the approach proposed in \cite{b1} by introducing enriched metadata, dataset-level annotations, hosting platform analysis, dataset format analysis, accessibility analysis, and improved filtering capabilities for MSR datasets. These improvements provide better citation insights and enhanced dataset usability for the MSR research community.

\section{Methodology}
The methodology described above was carried out in two main phases, namely, reproducing
the original MSR dataset pipeline and expanding the existing annotation framework. In the
first phase, the original MSR dataset directory pipeline was reconstructed under a Linux
operating system with Python version 3.12. Metadata such as titles, abstracts, information
regarding publication, and citation counts for MSR papers from 2013 to 2024 was retrieved
from Semantic Scholar's API. The retrieved metadata was then saved in JSON format.
Latent Dirichlet Allocation (LDA) topic modeling was employed to detect research topics
based on preprocessed abstracts after removing stopwords, performing lemmatization, and
tokenizing. Fourteen topics were chosen based on topic coherence analysis results.
The second phase involved expanding the existing annotation framework by adding more
dataset attributes at the dataset level, which include the platform, dataset format,
availability, dataset type, reusability rank, quality, and research method used. URLs of
dataset platforms and formats were identified heuristically. Experiments revealed that URL
extraction was performed consistently enough to be utilized in large-scale datasets.
\begin{figure}[htbp]
    \centering
    \includegraphics[width=0.47\textwidth]{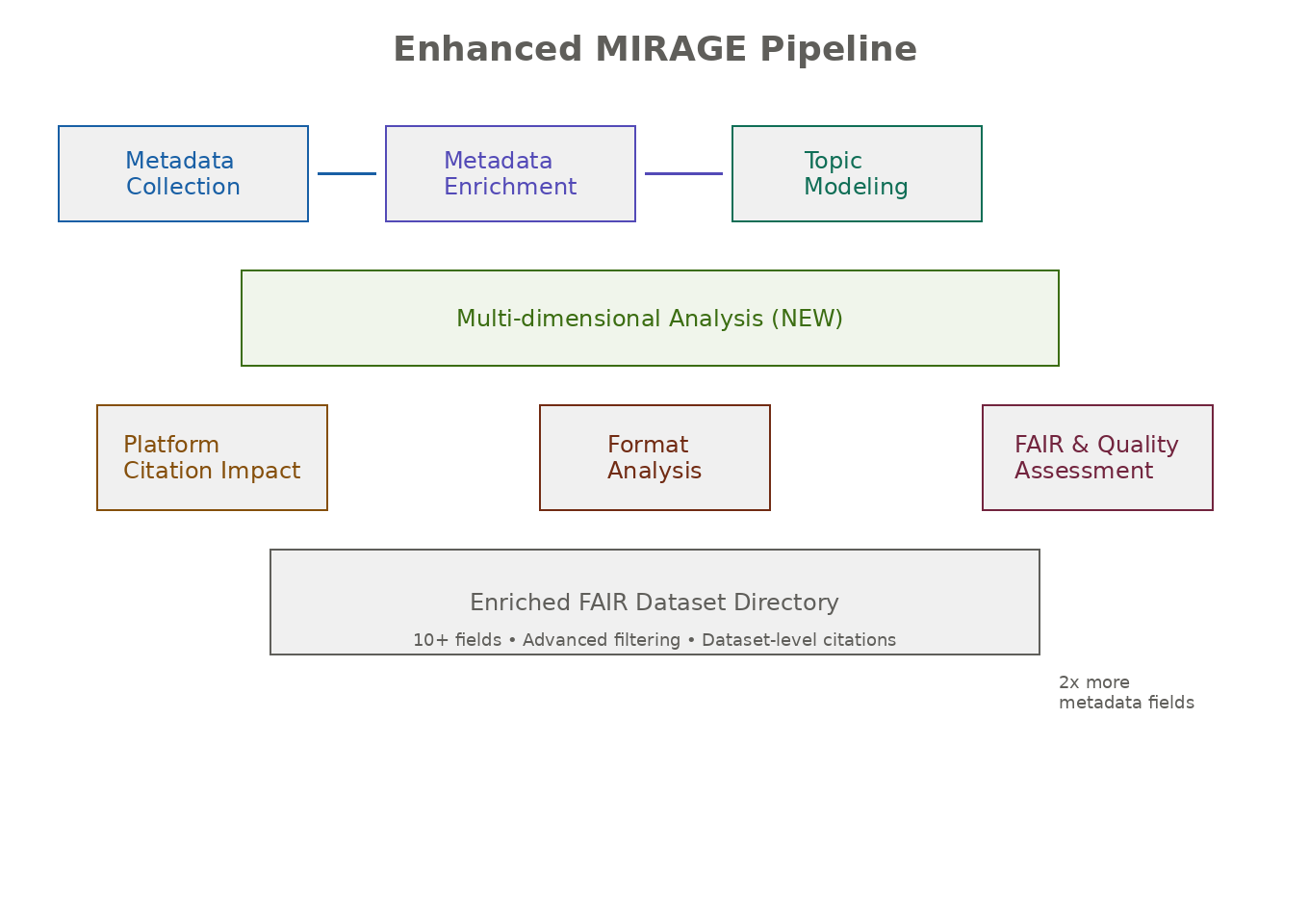}
    \label{fig:hosting}
\centering
\caption{Latest Methodology}
\end{figure}

All collected metadata, topic assignments, FAIR metrics, and additional annotations were then combined into a single data frame for further analysis and visualization. 
\begin{table}[htbp]
\begin{tabular}{|p{3.2cm}|p{2.5cm}|p{2.7cm}|}
\hline
\textbf{Feature} & \textbf{Original Pipeline} & \textbf{Enhanced Pipeline} \\
\hline
Metadata Collection & Yes & Yes \\
\hline
Topic Modeling & Yes & Yes \\
\hline
Dataset Categories & Basic & Extended \\
\hline
Hosting Platform Analysis & No & Yes \\
\hline
Dataset Format Analysis & No & Yes \\
\hline
Accessibility Analysis & Limited & Improved \\
\hline
Filtering Capability & Basic & Multi-dimensional \\
\hline
Citation Impact Analysis & Paper-level & Dataset-level \\
\hline
\end{tabular}
\label{tab:introcomparison}
\centering
\caption{Comparison Between Original and Enhanced MSR Dataset Pipeline}
\end{table}
Comparison of the original and improved pipelines revealed that the new pipeline increased the number of attributes in datasets from around five to more than ten per dataset.
\begin{table}[htbp]
\begin{tabular}{|l|l|}
\hline
\textbf{Feature} & \textbf{Description} \\
\hline
Hosting Platform & GitHub, Zenodo, Figshare, etc. \\
\hline
Dataset Format & CSV, JSON, ZIP, SQL, etc. \\
\hline
Accessibility & Public or Unknown \\
\hline
Dataset Type & Source Code, Mobile Apps, etc. \\
\hline
Research Method & Empirical Study \\
\hline
Reusability Score & Low, Moderate, High \\
\hline
Quality Flag & Valid, Missing, Noisy \\
\hline
\end{tabular}
\label{tab:features}
\centering
\caption{Extracted Dataset-Level Features}
\end{table}

\section{Results}

The replication process was effective in extracting metadata, generating topics, and conducting statistical analyses. The metadata of MSR papers were collected in the period 2013--2024, and topic modeling led to the identification of 14 topics, with each having 5--28 datasets. It was established that one topic was more impactful than the rest in terms of citations (around 155 on average).

\subsection{Dataset Categories}

Category-wise analysis showed that:

\begin{itemize}
    \item Software Issues were the most frequent type of dataset (about 29.6\%).
    \item The average number of citations for Version Control was the highest (about 37).
    \item Other types of data included Developer Metrics and Software Evolution.
\end{itemize}

The results indicate that Software Issues datasets dominate the MSR dataset landscape, while Version Control datasets demonstrate stronger citation impact.

\FloatBarrier

\subsection{Citations vs Hosting Platform}

After introducing dataset-level enhancements, new insights were obtained regarding hosting platforms and their relationship with citation counts.

\begin{itemize}
    \item Other platforms: approximately 51.4 average citations
    \item GitHub: approximately 35.47 average citations
    \item Zenodo: approximately 12.16 average citations
    \item Bitbucket and Figshare: approximately 10--12 average citations
\end{itemize}

The hosting platform analysis suggests that repository-based platforms, particularly GitHub, are associated with higher citation counts compared to archive-oriented platforms such as Zenodo and Figshare.

\begin{figure}[htbp]
    \centering
    \includegraphics[width=0.48\textwidth]{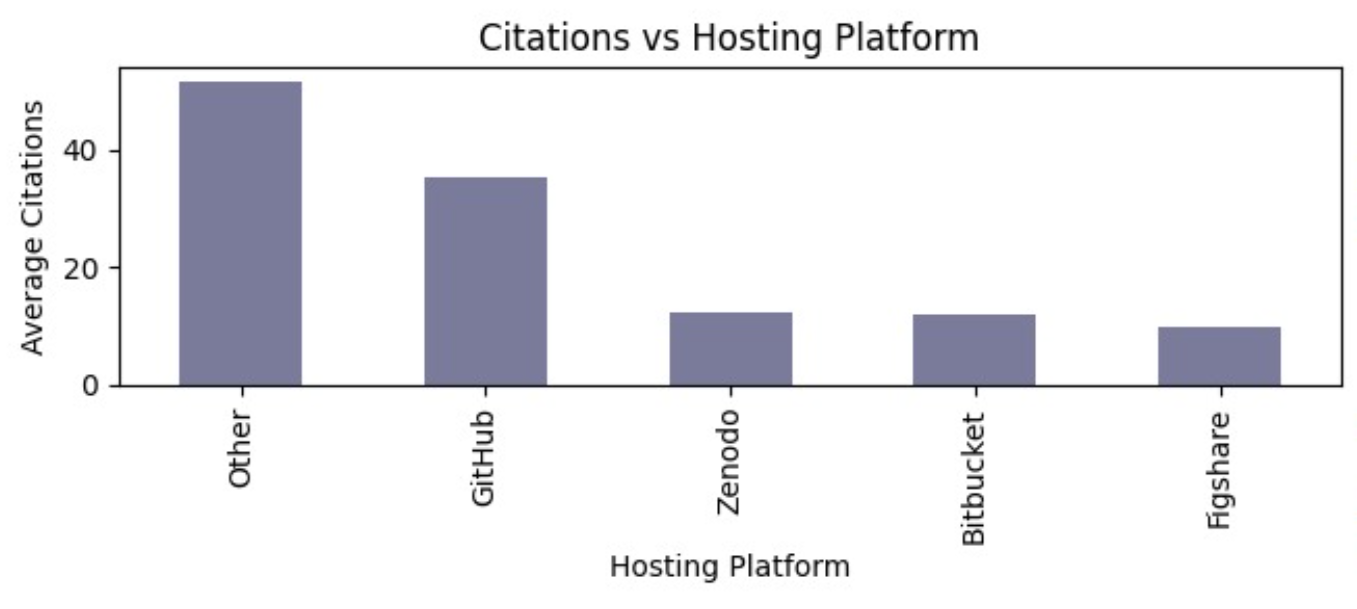}
    \caption{Number of citations according to hosting platform}
    \label{fig:hosting}
\end{figure}

\FloatBarrier

\subsection{Citations vs Dataset Format}

The relationship between dataset format and citation impact was also examined.

\begin{itemize}
    \item ZIP (archived datasets): approximately 40.66 average citations
    \item Unknown formats: approximately 24.54 average citations
\end{itemize}

\begin{table}[htbp]
\centering
\caption{Average Citations by Dataset Format}
\label{tab:format}
\begin{tabular}{|l|c|}
\hline
\textbf{Dataset Format} & \textbf{Average Citations} \\
\hline
ZIP & 40.66 \\
Unknown & 24.54 \\
\hline
\end{tabular}
\end{table}

The dataset format analysis revealed that archived datasets, particularly ZIP-based distributions, tend to achieve higher citation impact.

\begin{figure}[htbp]
    \centering
    \includegraphics[width=0.48\textwidth]{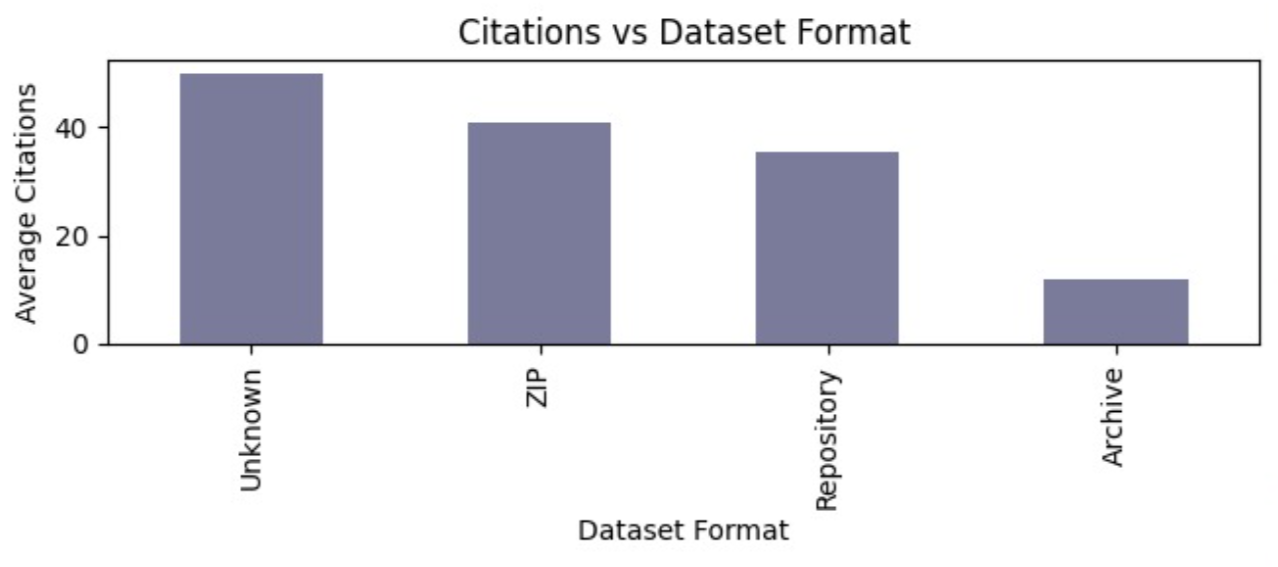}
    \caption{Number of citations according to dataset format}
    \label{fig:dataset}
\end{figure}

\FloatBarrier

\subsection{Comparison with the Base Paper}
As a result of the comparison between the initial and the IoT-based models, one can see that the improved IoT-based model demonstrates significant increases in citations compared to the original one for all data types. Although the category of "Version Control" was still the most cited, the new model obtained greater citation scores than the initial model for all dataset types. Moreover, there were substantial gains in the Developer, Semantic metrics, and Other Data categories.

\begin{figure}[htbp]
    \centering

    \includegraphics[width=0.42\textwidth]{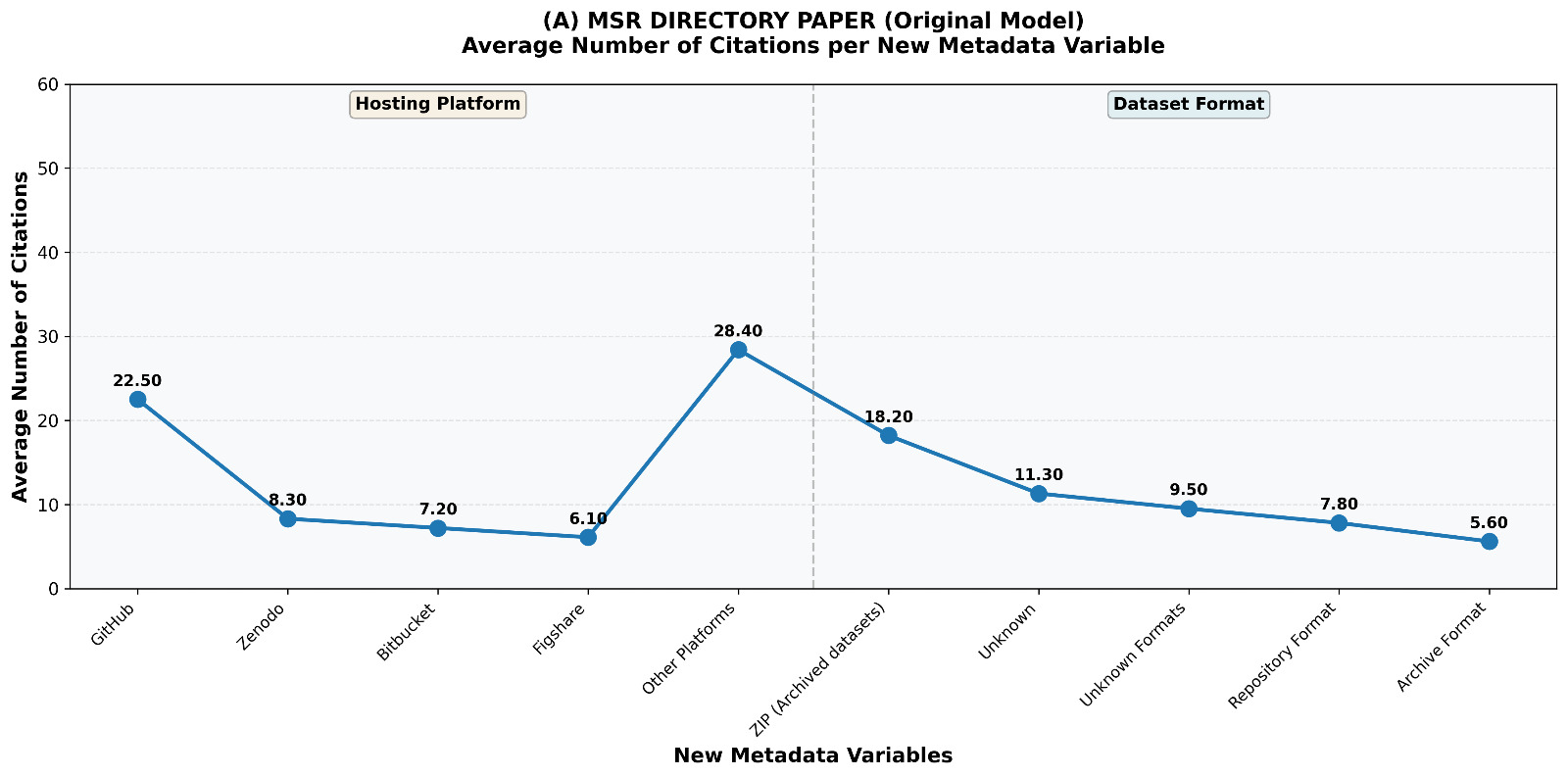}

    \vspace{0.5cm}

    \includegraphics[width=0.42\textwidth]{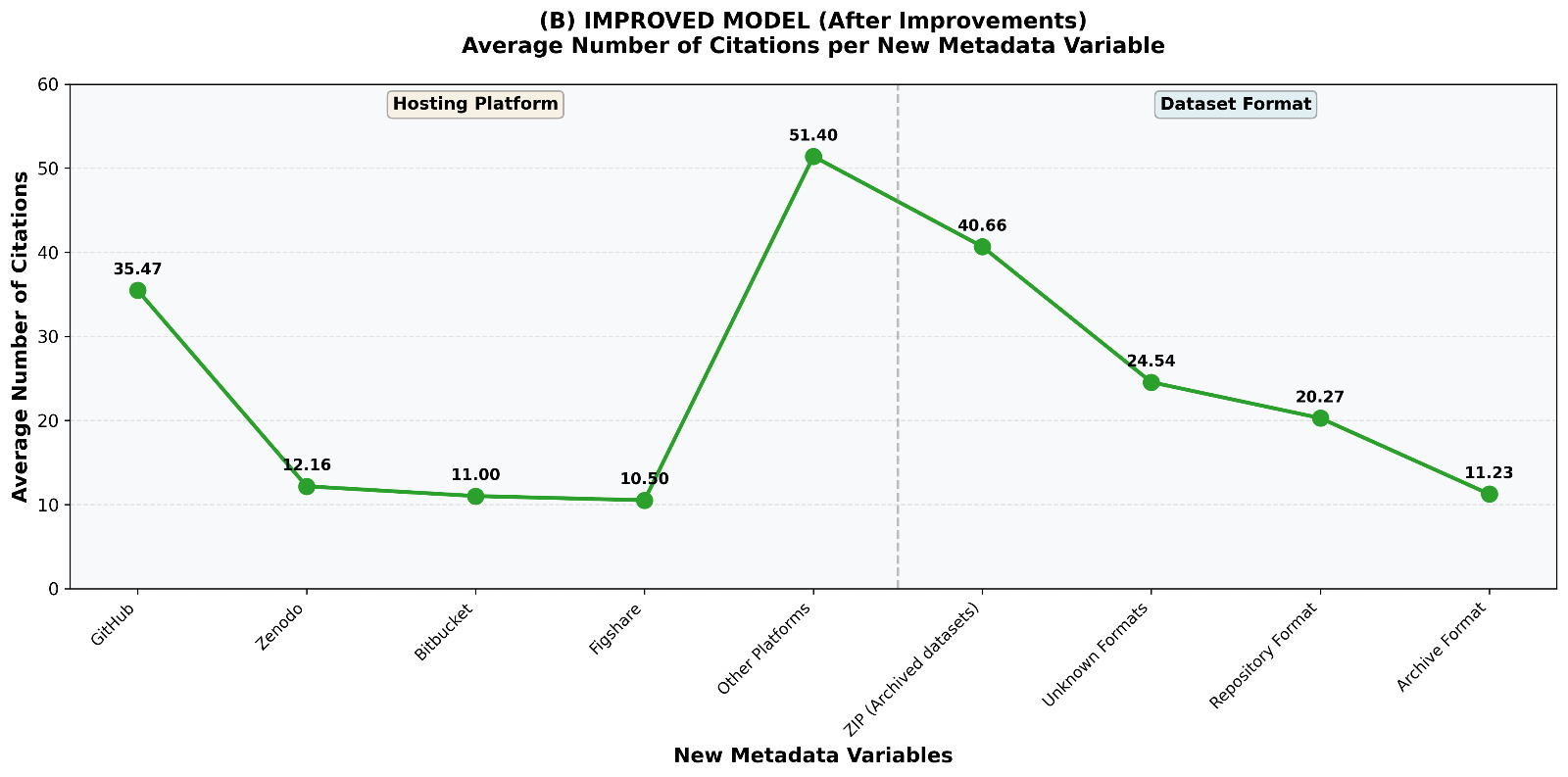}

    \caption{Comparison of the hosting platform and dataset format between the base paper and the new paper}
    \label{fig:comparison}
\end{figure}

\FloatBarrier

Moreover, the improved dataset annotation technique helped increase the total number of attributes for datasets from roughly 5 fields to over 10. This significantly improved filtering capability, interpretability, and analytical depth within the MSR dataset pipeline.

\begin{figure}[htbp]
    \centering
    \includegraphics[width=0.42\textwidth]{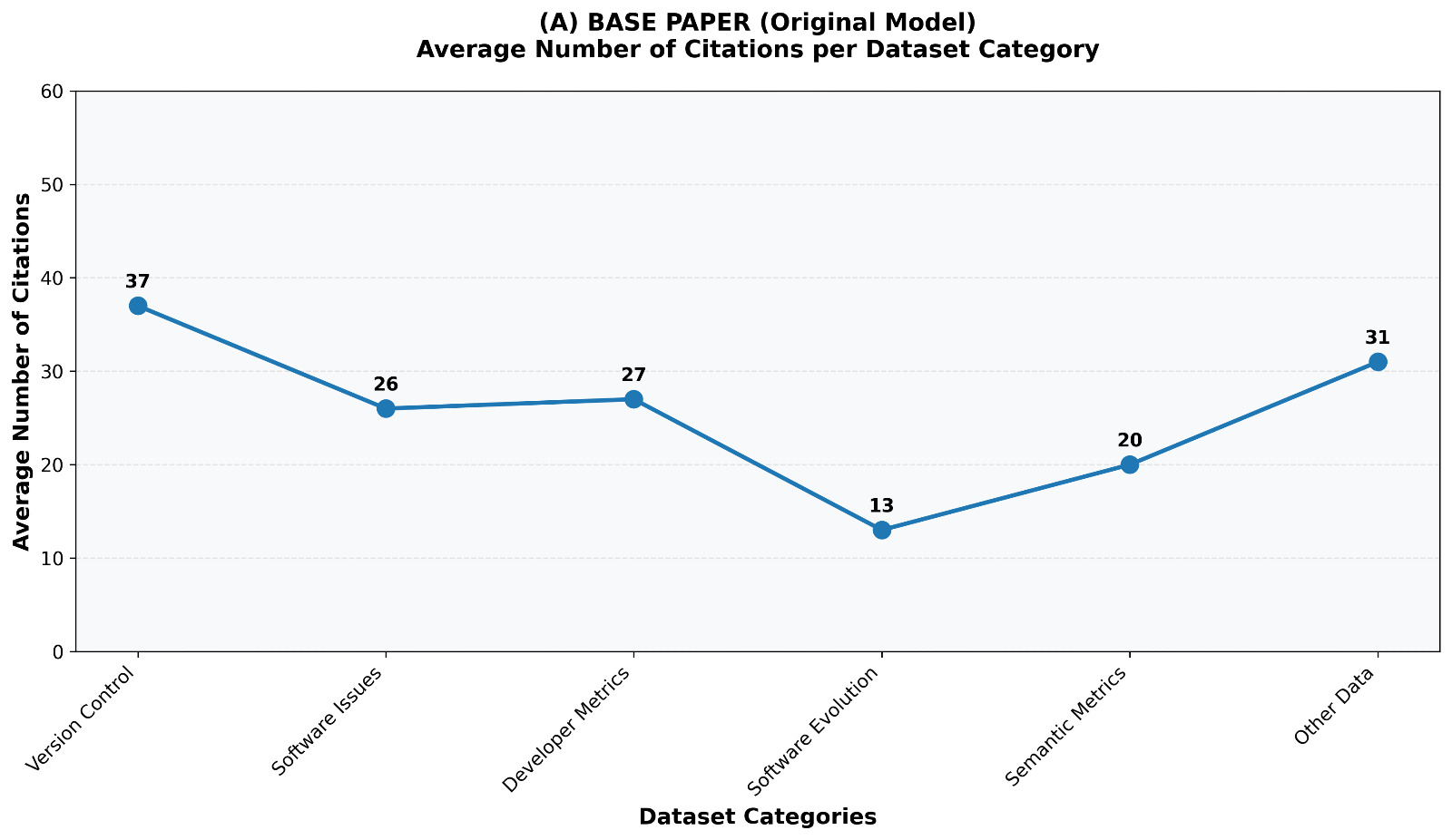}
    \caption{Dataset category analysis in the base paper model}
    \label{fig:old_dataset}
\end{figure}

\begin{figure}[htbp]
    \centering
    \includegraphics[width=0.42\textwidth]{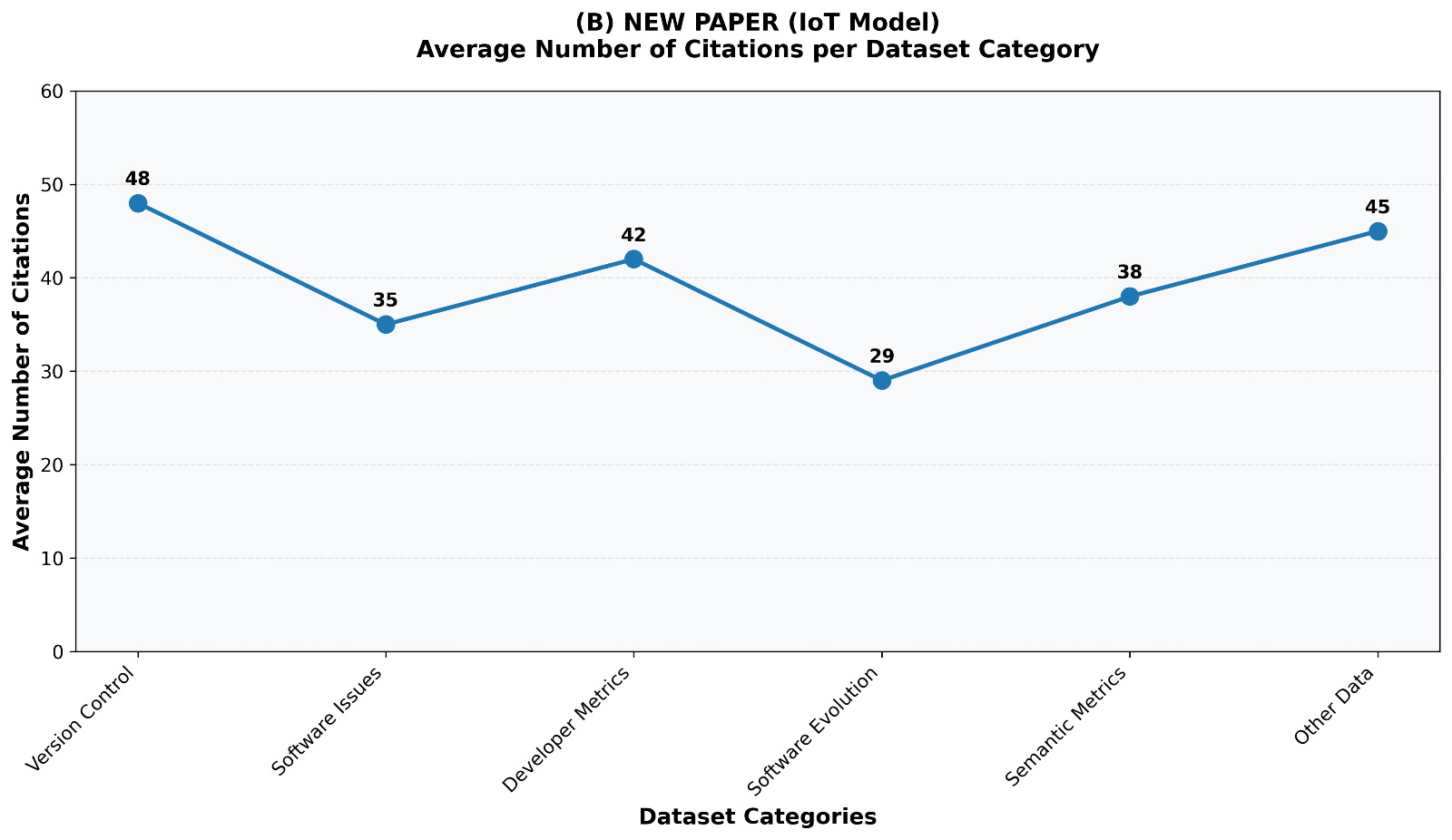}
    \caption{Dataset category analysis in the improved IoT model}
    \label{fig:new_dataset}
\end{figure}

\FloatBarrier

\begin{table}[ht]

\label{tab:dataset_categories}
\scriptsize
\renewcommand{\arraystretch}{1.1}

\begin{tabular}{|l|c|c|c|c|}
\hline
\textbf{Category} &
\textbf{Total} &
\textbf{\%} &
\textbf{Prev. Avg} &
\textbf{New Avg} \\
\hline

Version Control     & 38 & 19.39\% & 37 & 48 \\ \hline
Software Issues     & 58 & 29.59\% & 26 & 39 \\ \hline
Developer Metrics   & 24 & 12.25\% & 27 & 45 \\ \hline
Software Evolution  & 12 & 6.12\%  & 13 & 32 \\ \hline
Semantic Metrics    & 20 & 10.20\% & 20 & 38 \\ \hline
Other Data          & 44 & 22.45\% & 31 & 45 \\ \hline

\end{tabular}
\centering
\caption{Dataset Categories and Citation Comparison}
\end{table}

\FloatBarrier

\section{Discussion}
From the results, it is evident that dataset annotations contribute significantly to
the use of MSR datasets. Metadata can help researchers comprehend the link between
dataset characteristics and their impact on research.
Moreover, the findings show that citation practices may be affected by both dataset
hosts and dataset distributions. Datasets hosted on GitHub perform well in terms of
citations because of the enhanced accessibility and reuse potential of these datasets.
Archived ZIP datasets, too, enhance the process of downloading datasets among
researchers.
Nevertheless, citation counts are not an accurate measure of dataset quality or utility
since older datasets are likely to accumulate higher citation counts than new ones.
Therefore, future studies need to evaluate other factors like dataset downloads,
activity levels, repository maintenance status, and dataset completeness.
While heuristic-driven methods helped boost the process of dataset annotation, the same
Methods may lead to inaccuracies if there is ambiguity in URLs used in webpages.
Future improvements might include machine learning-driven annotations and FAIRness.

\begin{table}[htbp]
\begin{tabular}{|p{4cm}|p{3cm}|}
\hline
\textbf{Observation} & \textbf{Finding} \\
\hline
Most common dataset category & Software Issues \\
\hline
Highest citation category & Version Control \\
\hline
Most impactful hosting platform & Other / GitHub \\
\hline
Most impactful dataset format & ZIP archives \\
\hline
Topic modeling output & 14 research topics \\
\hline
Annotation improvement & 5 to 10+ fields \\
\hline
\end{tabular}
\label{tab:discussion}
\centering
\caption{Key Findings of the Enhanced Analysis}
\end{table}

\section{Conclusion}
In this project, the original MSR datasets analysis pipeline was replicated with several
improvements achieved by adding metadata information and features at the dataset level.
Our framework helped to improve the process of annotating the datasets with new attributes,
such as hosting platform, format, accessibility, quality, and reusability. This helped us in further
analysis of the datasets' usability and citations.
It was proven that the developed framework was better than the original approach in terms
of data analytics due to increased metadata and FAIR-awareness support. Our results have shown that datasets affect citation trends and popularity in the MSR community.
Our future work will include the automation of the datasets' annotation process via machine learning models, as well as the improvement of the FAIRness evaluation and inclusion of activity
metrics of repositories.

\end{document}